\documentclass[sigconf]{acmart}

\AtBeginDocument{%
  \providecommand\BibTeX{{%
    \normalfont B\kern-0.5em{\scshape i\kern-0.25em b}\kern-0.8em\TeX}}}

\copyrightyear{2019}
\acmYear{2019}
\setcopyright{rightsretained}

\copyrightyear{2019}
\acmYear{2019}
\acmConference[REVEAL '19]{REVEAL '19}{September 16--20, 2019}{Copenhagen, Denmark}
\acmPrice{15.00}
\acmDOI{xx.xxxx/xxxxxxx.xxxxxxx}
\acmISBN{xxx-x-xxxx-xxxx-x/xx/xx}

\usepackage{algorithm}   % REMOVE
\usepackage[noend]{algpseudocode} %REMOVE
\usepackage{graphics}
\usepackage{subcaption}

\DeclareMathOperator*{\diversity}{div}
\DeclareMathOperator*{\dist}{dist}
\DeclareMathOperator*{\anDCG}{\alpha-nDCG}
\DeclareMathOperator*{\aIDCG}{aIDCG}
\DeclareMathOperator*{\rank}{rank}
\DeclareMathOperator*{\rel}{rel}
\DeclareMathOperator*{\ILD}{ILD}
\DeclareMathOperator*{\SD}{SD}
\DeclareMathOperator*{\hit}{hit}
\DeclareMathOperator*{\RL}{RL}

\begin{document}

\title{Sudden Death:\\A New Way to Compare Recommendation Diversification}

%
% The "author" command and its associated commands are used to define the authors and their affiliations.
% Of note is the shared affiliation of the first two authors, and the "authornote" and "authornotemark" commands
% used to denote shared contribution to the research.
\author{Derek Bridge}
\email{derek.bridge@insight-centre.org}
\affiliation{%
  \institution{Insight Centre for Data Analytics}
  %\streetaddress{P.O. Box 1212}
  \city{University College Cork}
  \state{Ireland}
}

\author{Mesut Kaya}
\email{mesut.kaya@insight-centre.org}
\affiliation{%
  \institution{Insight Centre for Data Analytics}
  %\streetaddress{P.O. Box 1212}
  \city{University College Cork}
  \state{Ireland}
}

\author{Pablo Castells}
\email{pablo.castells@uam.es}
\affiliation{%
  \institution{Universidad Aut\'{o}nonoma de Madrid}
  %\streetaddress{P.O. Box 1212}
  \city{Madrid}
  \state{Spain}
}

\renewcommand{\shortauthors}{Bridge et al.}
\renewcommand{\shorttitle}{Sudden Death}
\begin{abstract}
This paper describes problems with the current way we compare the diversity of different recommendation lists in offline experiments. We illustrate the problems with a case study. We propose the Sudden Death score as a new and better way of making these comparisons.
\end{abstract}


\maketitle

\section{Introduction} \label{sec:intro}

Recommendations should not only be relevant; a set of recommendations should often also be diverse. A number of algorithms exist that diversify their top-$N$ recommendation lists; a number of metrics exist that measure diversity. 

But when \emph{comparing} algorithms, especially in offline evaluations, there are at least two problems. 

The first problem is that the comparison may not be fair. A recommender that seeks to diversify may use an objective function that is the same as, or closely related to, the metric used to measure diversity. This metric will tend to favour this recommender. 

The second problem is that we are usually not interested in diversity for its own sake, although this is what most of the diversity metrics measure. The recommendations in the top-$N$ must still be relevant to the user. Checking this in the results of offline experiments becomes more challenging because we now have a multi-objective evaluation. %There is often said to be a trade-off between relevance and diversity. This is strange when one considers that the main reason that we diversify is to make it more likely that the user will find an item that satisfies her. There should be no trade-off: if diversification is working as it should, users should be more likely to find a relevant item. 

In this short paper, we propose a new scoring method for \emph{comparing} algorithms. It is not specific to diversity but it is inspired by the idea that diversification is supposed to make it more likely that the user finds an item that satisfies her. 

In Section \ref{sec:problems}, we illustrate the problems that we have presented above. In Section \ref{sec:sd}, we give a definition of the Sudden Death score and we illustrate its use. In Section \ref{sec:conc}, we make concluding remarks.

\section{A Case Study} \label{sec:problems}

We illustrate the two problems by comparing four algorithms using three diversity metrics and three relevance metrics on a single dataset. Of course, if our goal were to find the `best' algorithm, we would use more datasets. But our purpose is simply to motivate the need for a better way of comparing algorithms. 

We use the MovieLens 1 Million dataset.\footnote{https://grouplens.org/datasets/movielens/1m/} 

The algorithms we compare are MMR, xQuAD and SPAD. They greedily re-rank the recommendations made by a baseline recommender, for which we use  Matrix Factorization (MF). In these algorithms, the baseline recommender produces a set of recommended items, $RS$, for user $u$. For each recommended item $i$ in $RS$, it also produces a score, $s(u, i)$, that estimates the relevance of recommended item $i$ to user $u$. Then, the greedy algorithm re-ranks $RS$ by iteratively inserting into ordered result list $RL$ the item $i$ from $RS$ that maximizes a function, $f_{\mathit{obj}}(i, RL)$:
\begin{equation}\label{eq:greedyobj}
f_{\mathit{obj}}(i, RL) = (1 - \lambda) s(u,i) + \lambda \diversity(i, RL)
\end{equation}
where $\diversity(i, RL)$ is the marginal gain in diversity after inserting item $i$ into the list of already-selected items $RL$. In this objective function, the trade-off between relevance and diversity is controlled by a parameter $\lambda$ ($0 \leq \lambda \leq 1$). MMR, xQuAD and SPAD differ in their definitions of diversity, $\diversity(i, RL)$. 

In MMR, $\diversity(i,RL)$ is the maximum of the distances between $i$ and the items already selected \cite{carbonell1998use}:
\begin{equation}
    \diversity(i,RL)= \max_{j \in RL}\dist(i,j)
\end{equation}
The distance between items $i$ and $j$, $\dist(i,j)$, is most often calculated from item features (such as movie genres or book categories).

xQuAD and SPAD are intent-aware diversification methods \cite{Vargas-2015}. They try to achieve coverage of the user's different interests, as revealed by the user's profile. More specifically, a user $u$'s interests are formulated as a probability distribution $p(a|u)$ over a set of aspects $a \in \mathcal{A}$. Then, diversity is defined as follows:
\begin{equation} \label{eq:IAD}
\diversity(i, RL) = \sum_{a \in \mathcal{A}} [p(a|u)p(i|u,a)\prod_{j \in RL}(1 - p(j|u,a))]
\end{equation}
where $p(i|u, a)$ is the probability of choosing an item $i$ from a set of candidate recommendations $RS$, produced by a conventional recommender algorithm, given an aspect $a$ and user $u$. 

In xQuAD, the aspects are given by item features \cite{Vargas-2015}. In SPAD, aspects are subprofiles, which are mined from the items that the user likes \cite{Kaya-Bridge-2019}.

In the results shown in the rest of this paper, the methodology we adopt for training, for testing and for the setting of hyperparameter values is the same as the one described in \cite{Kaya-Bridge-2019}.

\begin{figure*}[t]
    \centering
    \begin{subfigure}[b]{0.3\textwidth}
        \includegraphics[width=\textwidth]{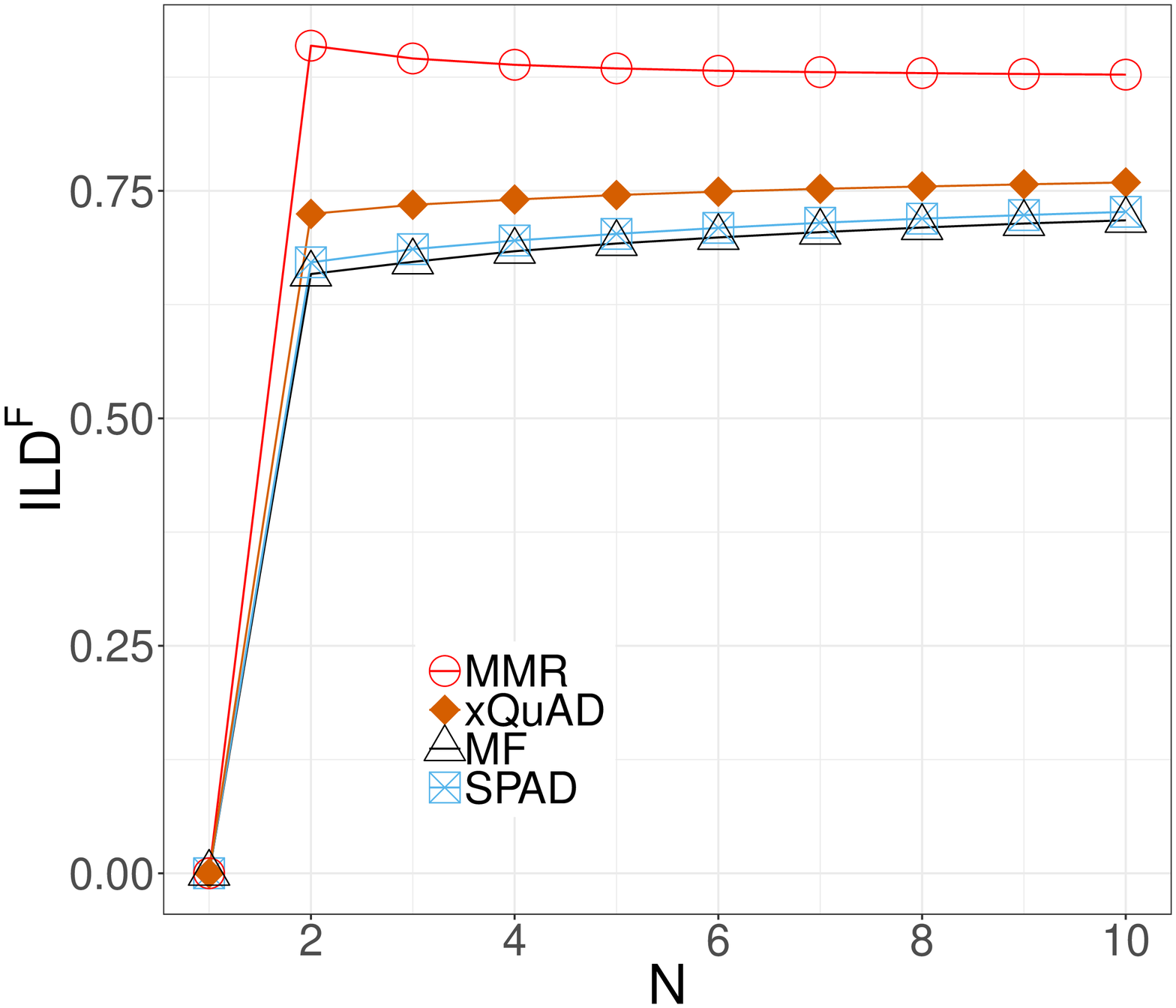}
        \caption{$\ILD$}
        \label{fig:div-ild}
    \end{subfigure}
        ~ %add desired spacing between images, e. g. ~, \quad, \qquad, \hfill etc. 
    \begin{subfigure}[b]{0.3\textwidth}
        \includegraphics[width=\textwidth]{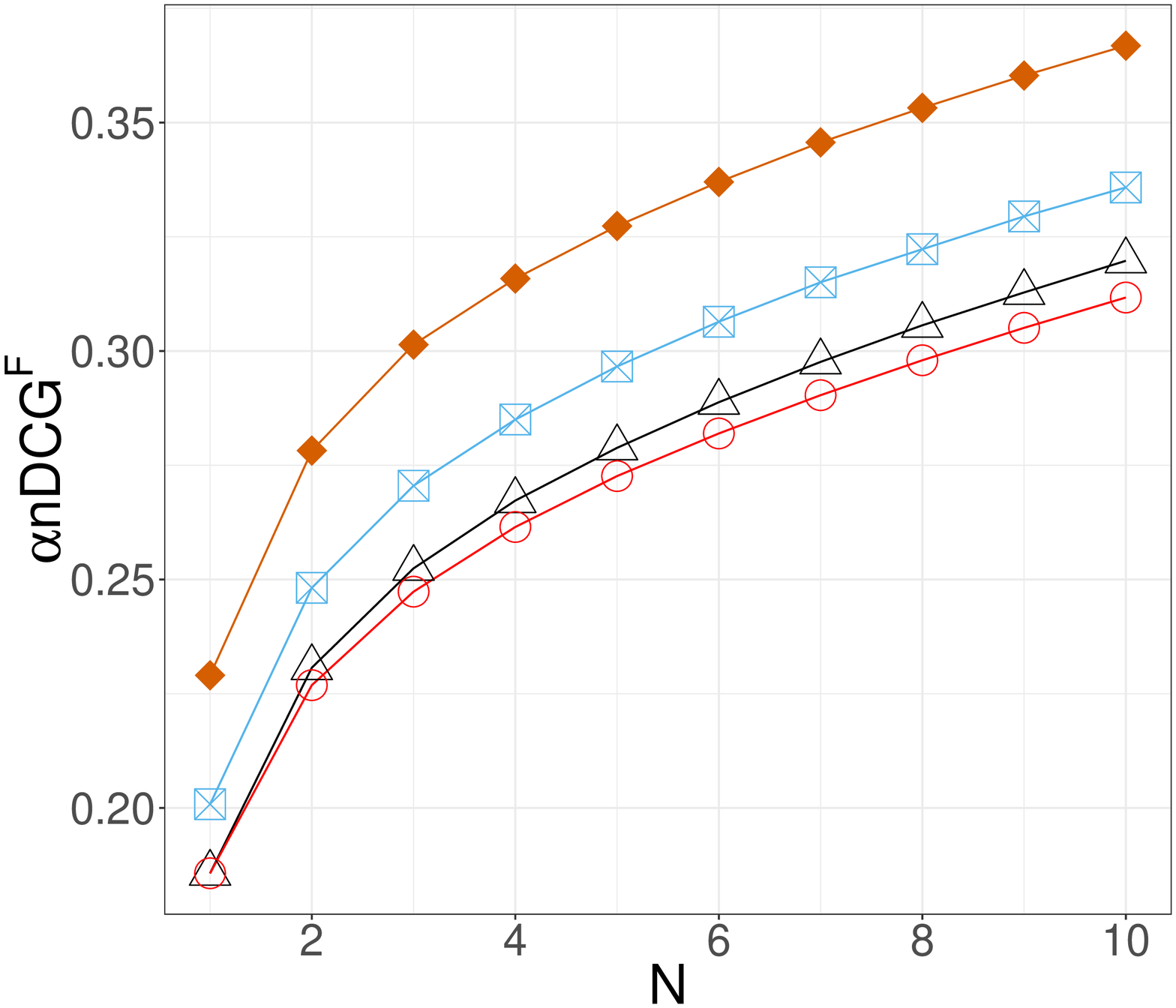}
        \caption{$\anDCG^{\mathcal{F}}$}
        \label{fig:div-andcg-f}
    \end{subfigure}
            ~ %add desired spacing between images, e. g. ~, \quad, \qquad, \hfill etc. 
    \begin{subfigure}[b]{0.3\textwidth}
        \includegraphics[width=\textwidth]{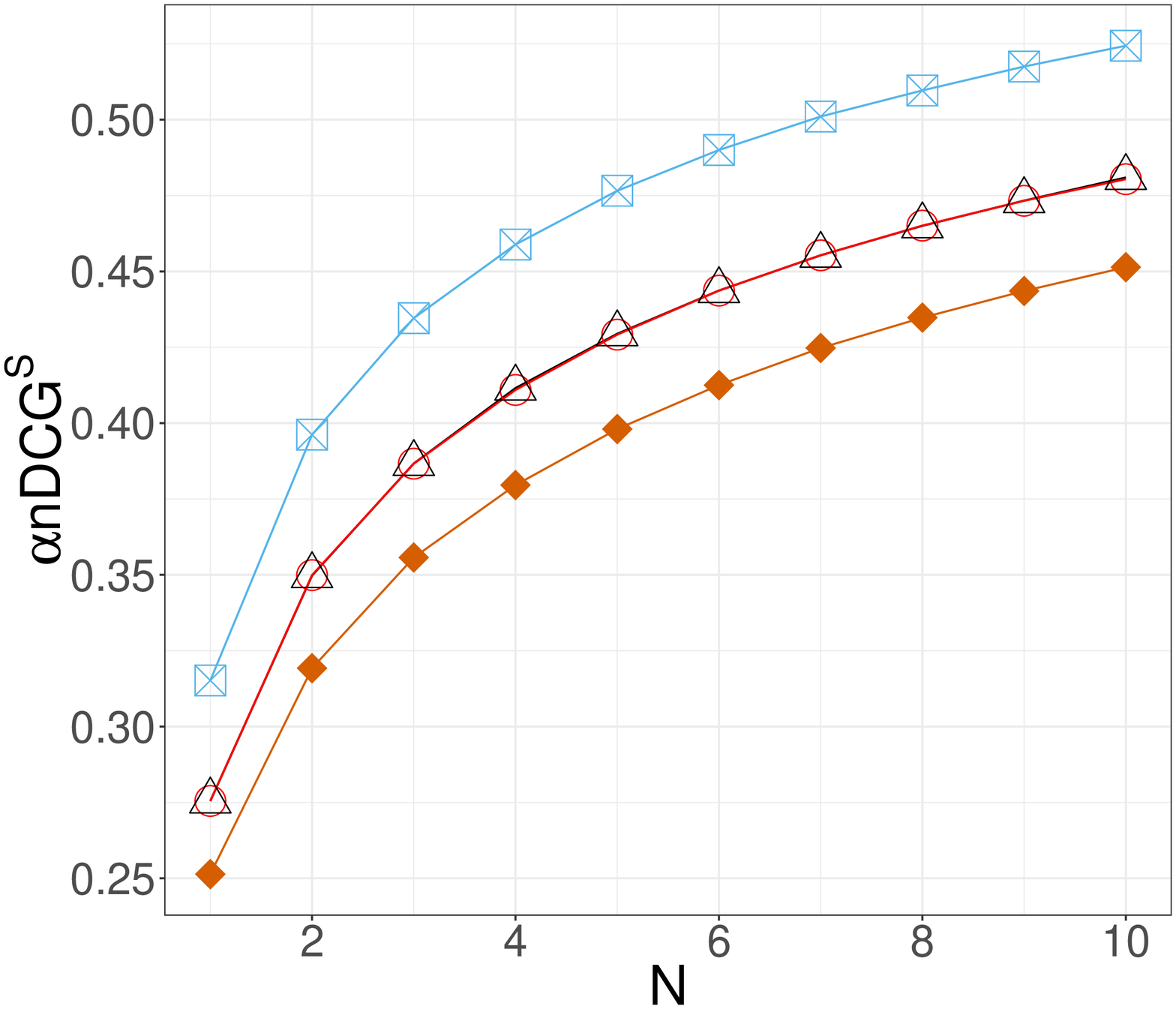}
        \caption{$\anDCG^{\mathcal{S}}$}
        \label{fig:div-andcg-s}
    \end{subfigure}
    \caption{Different diversity metrics for different values of $N$.}\label{fig:div-plots}
\end{figure*}

\subsection{Measuring diversity}

\begin{figure*}[t]
    \centering
    \begin{subfigure}[b]{0.3\textwidth}
        \includegraphics[width=\textwidth]{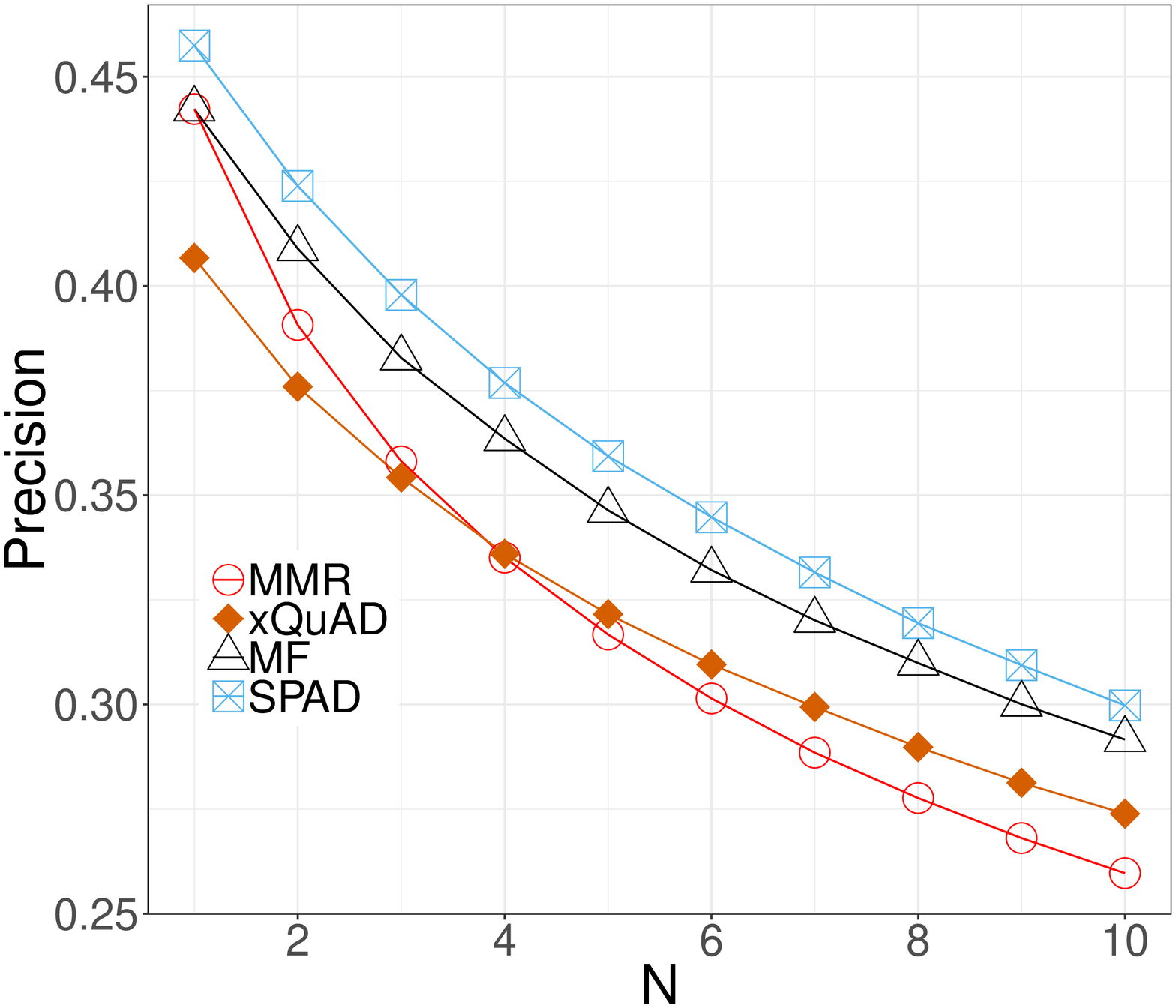}
        \caption{Precision}
        \label{fig:rel-p}
    \end{subfigure}
        ~ %add desired spacing between images, e. g. ~, \quad, \qquad, \hfill etc. 
    \begin{subfigure}[b]{0.3\textwidth}
        \includegraphics[width=\textwidth]{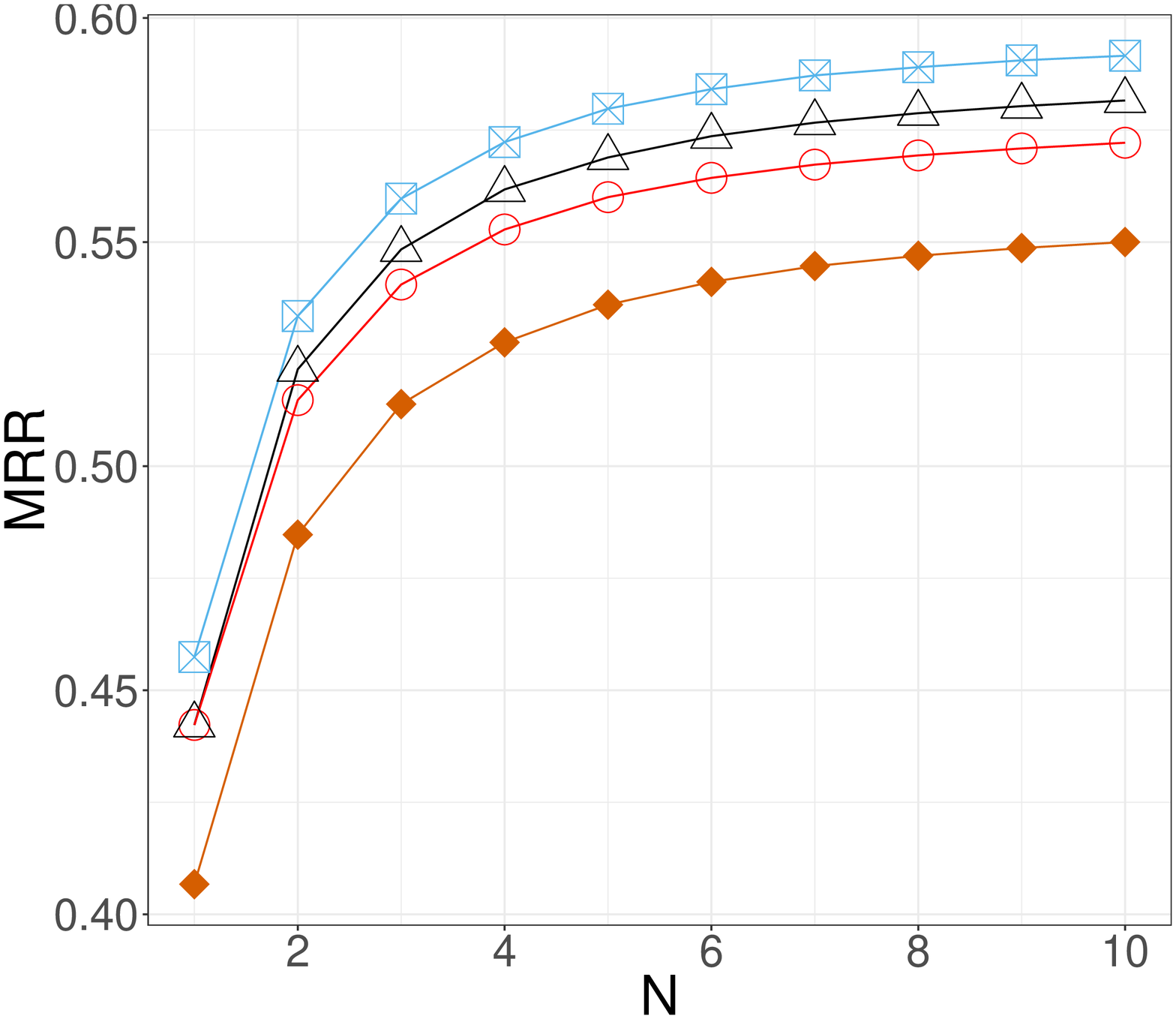}
        \caption{MRR}
        \label{fig:rel-mrr}
    \end{subfigure}
            ~ %add desired spacing between images, e. g. ~, \quad, \qquad, \hfill etc. 
    \begin{subfigure}[b]{0.3\textwidth}
        \includegraphics[width=\linewidth]{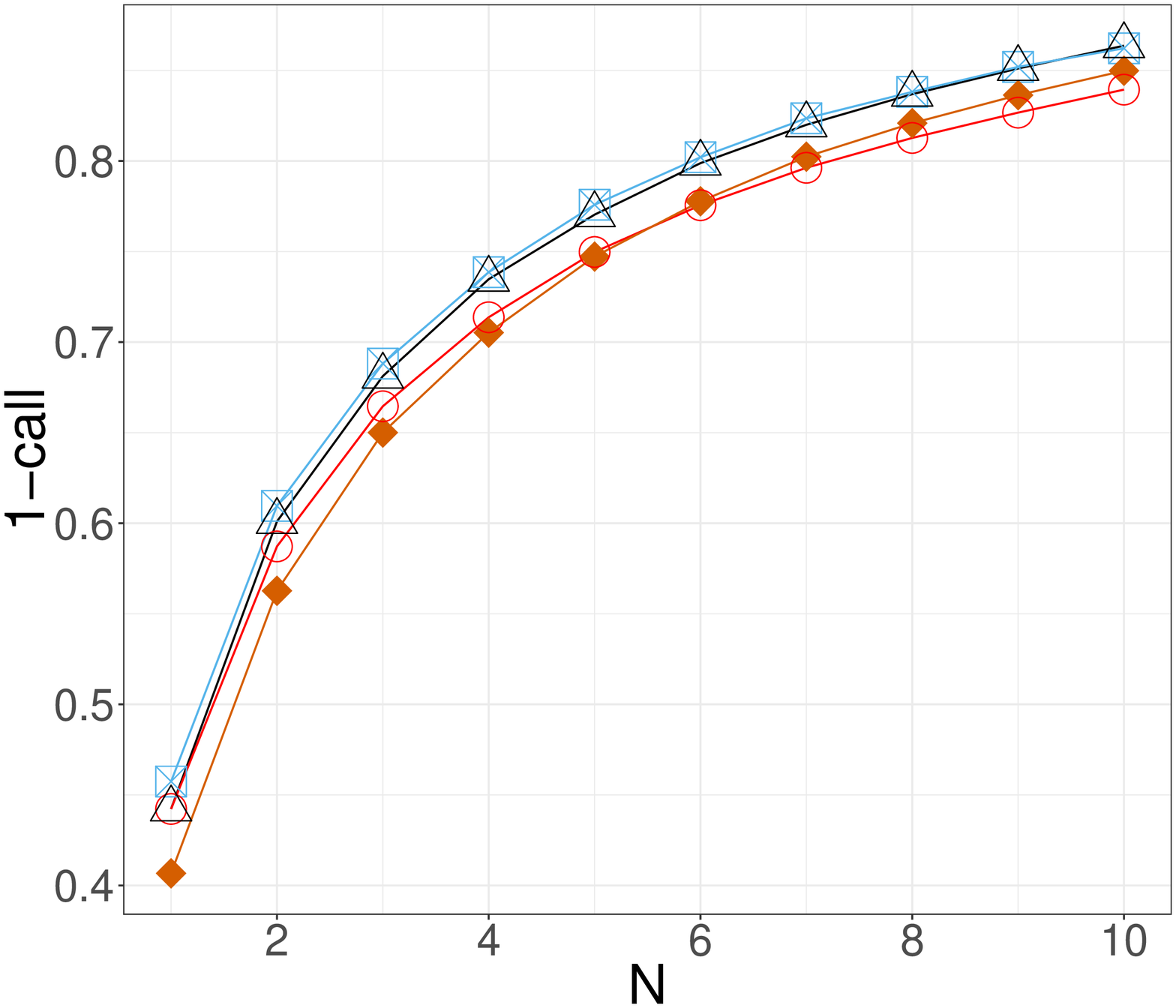}
        \caption{1-call}
        \label{fig:1call}
     \end{subfigure}
    \caption{Different relevance metrics for different values of $N$.}\label{fig:rel-plots}
\end{figure*}      

\begin{figure*}[t]
    \centering
    \begin{subfigure}[b]{0.3\textwidth}
        \includegraphics[width=\textwidth]{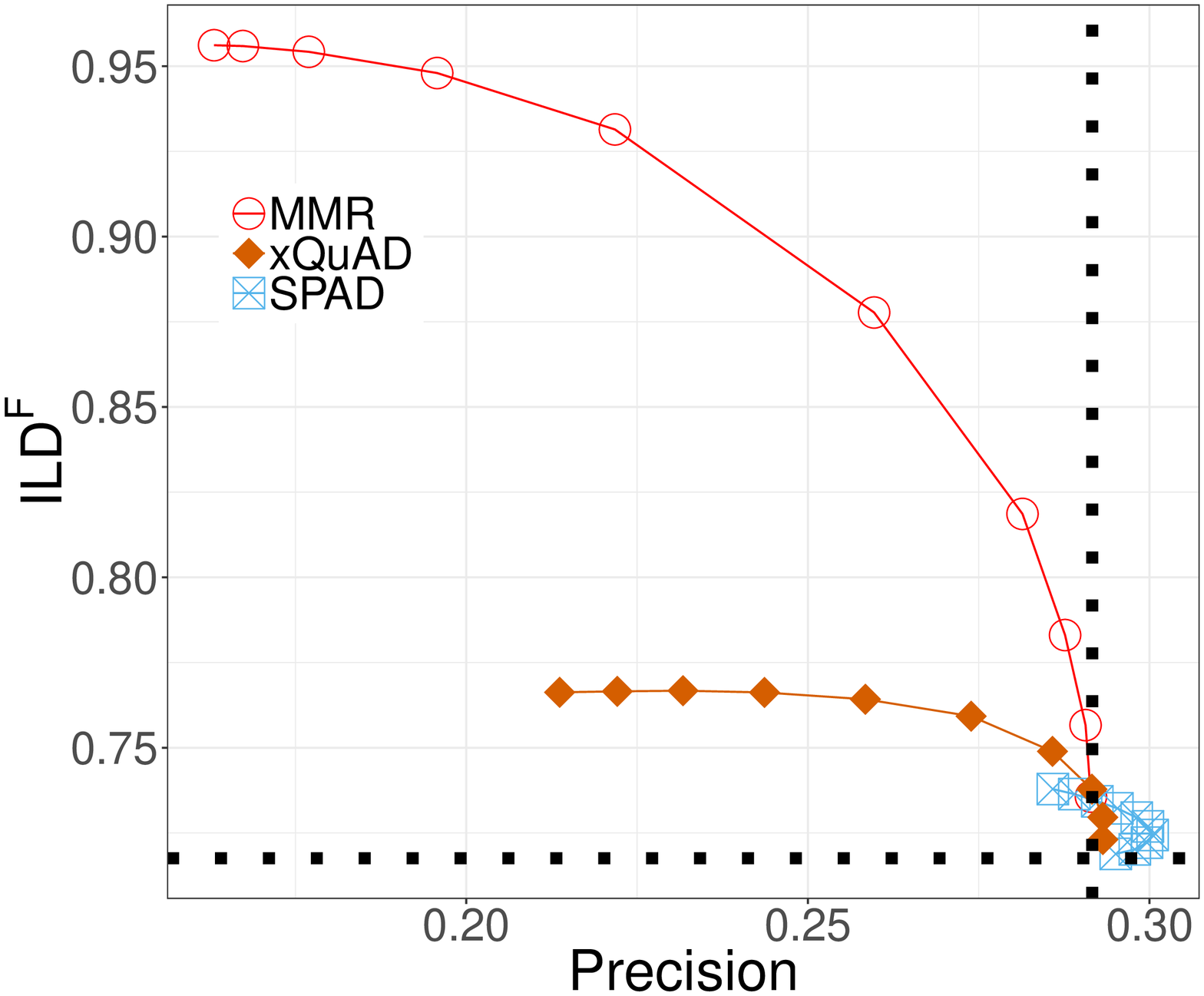}
        \caption{Prec./$\ILD^{\mathcal{F}}$}
        \label{fig:prec-ild-f}
    \end{subfigure}
        ~ %add desired spacing between images, e. g. ~, \quad, \qquad, \hfill etc. 
    \begin{subfigure}[b]{0.3\textwidth}
        \includegraphics[width=\textwidth]{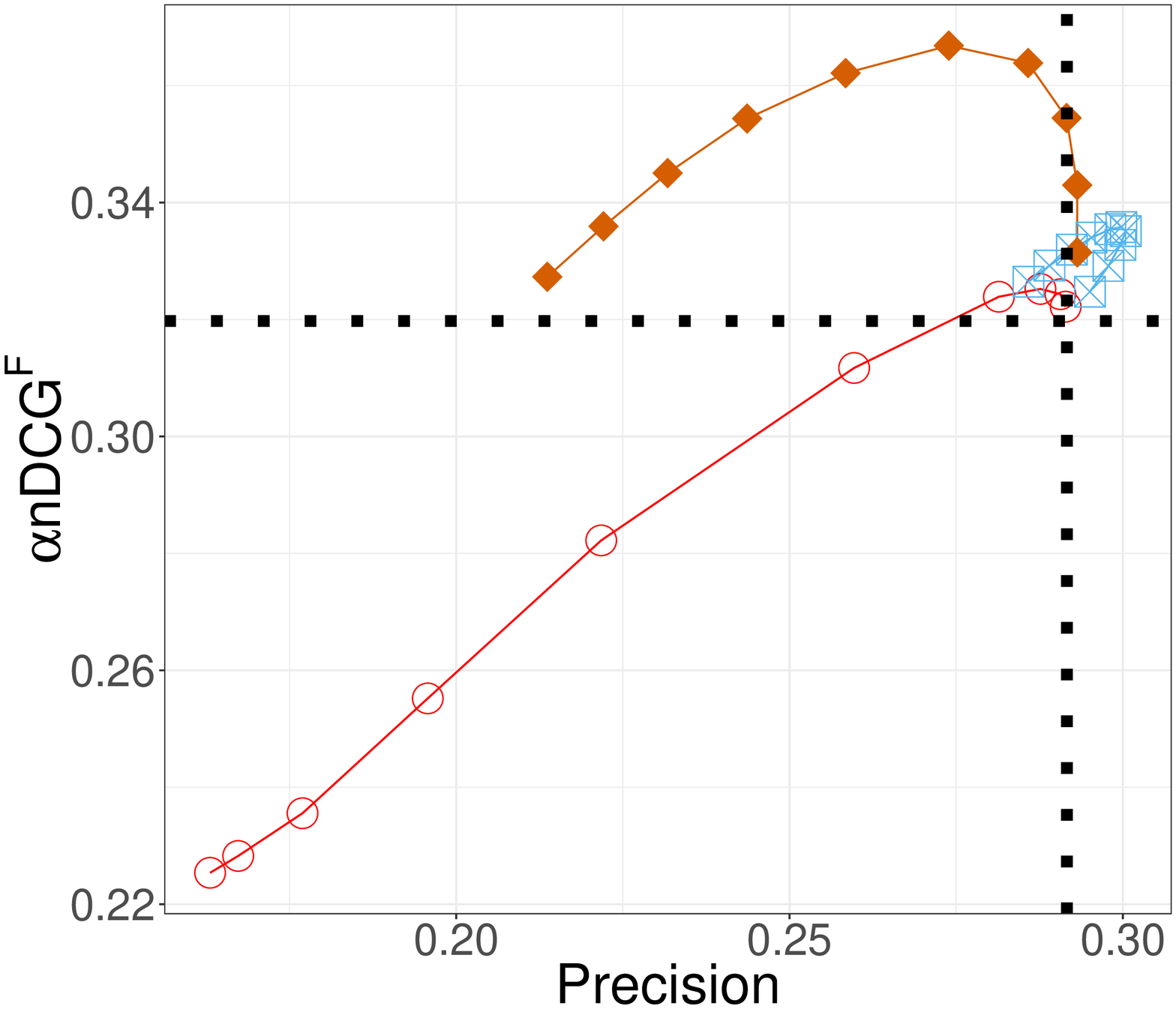}
        \caption{Prec./$\anDCG^{\mathcal{F}}$}
        \label{fig:prec-andcg-f}
    \end{subfigure}
            ~ %add desired spacing between images, e. g. ~, \quad, \qquad, \hfill etc. 
    \begin{subfigure}[b]{0.3\textwidth}
        \includegraphics[width=\textwidth]{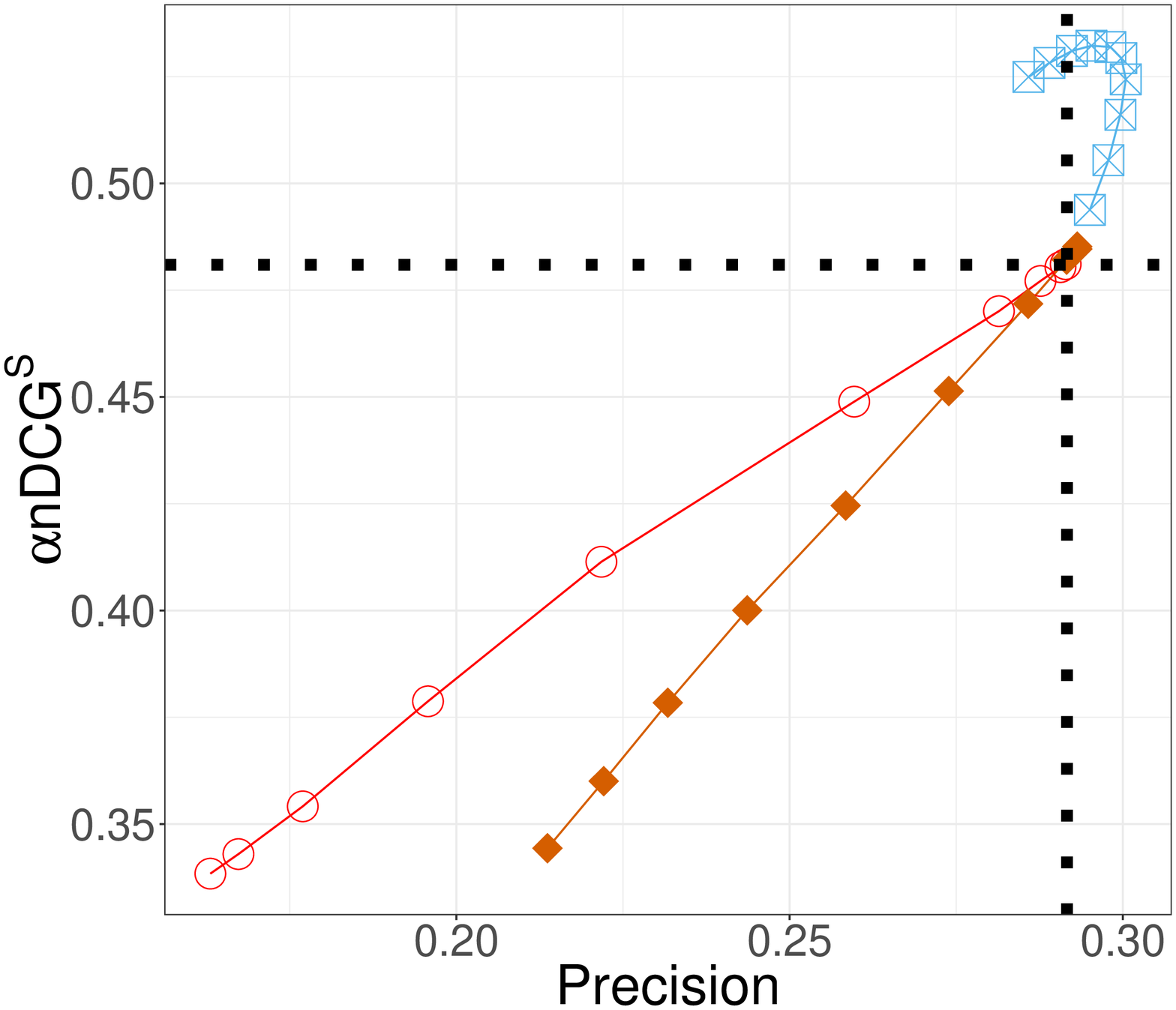}
        \caption{Prec./$\anDCG^{\mathcal{S}}$}
        \label{fig:prec-andcg-s}
     \end{subfigure}
    \caption{Precision vs.\ diversity for different values of $\lambda$. Dotted lines show the performance of the MF baseline.}\label{fig:tradeoff-plots}
\end{figure*}  

Just as there are several ways of measuring diversity within the objective function of a greedy re-ranking algorithm, there are several ways of measuring the diversity of a list of recommendations, $L$, for evaluation purposes.

Intra-List Diversity ($\ILD$) is a popular diversity metric; it computes the average pairwise distances of items in a recommendation set \cite{ziegler2005improving}:
\begin{equation} \label{eq:ILD}
\ILD(L) = \frac{2}{|L|(|L|-1)} \sum_{i \in L} \sum_{j \in L, j \neq i} \dist(i,j)
\end{equation}
$\ILD$ may favour an algorithm like MMR \cite{carbonell1998use}, since $\ILD$ is close to what MMR optimizes. This is confirmed in Figure \ref{fig:div-ild}.

$\anDCG$ is a newer diversity metric \cite{clarke2008novelty}. It measures coverage and relevance of aspects, $\mathcal{A}$:
\begin{multline}\label{eq:andcg}
\anDCG(L) = \frac{1}{\aIDCG} \sum_{i \in L} \left[\frac{1}{\log_2(\rank(i, L)+1)}\sum_{a \in \mathcal{A}}\right.\\ \left.\rel(i|u,a)  
 \prod_{\substack{j\in L,\\ \rank(j, L) < \rank(i, L)}}(1-\alpha \rel(j|u,a)) \right]
\end{multline}
%\end{equation}
where $\aIDCG$ is the highest possible value of $\anDCG$ where the recommendation set is made of ideally diversified relevant items, $L$ is the set of recommended items (of size $N$), $\rank(i, L)$ is the position of $i$ in $L$, and $\rel(i|u,a)$ is 1 if item $i$ has aspect $a$ and is relevant to user $u$ but 0 otherwise. $\alpha$ is the parameter that controls the penalty for redundancy. We use $\alpha = 0.5$, as in \cite{Vargas-2015}. Most commonly, aspects are item features. To make this explicit,, we will write $\anDCG^{\mathcal{F}}$. Clearly, this metric favours xQuAD; and this is confirmed in Figure \ref{fig:div-andcg-f}.

In \cite{Kaya-Bridge-2019b}, we modified $\anDCG$ to produce a variant that we will denote by $\anDCG^{\mathcal{S}}$, which defines aspects in terms of subprofiles. This metric favours SPAD, as confirmed by Figure \ref{fig:div-andcg-s}.

The three plots in Figure \ref{fig:div-plots} illustrate the first of the two problems that we presented in Section \ref{sec:intro}. Experiments that use just one  diversity metric are not fair because each metric tends to favour certain recommenders. 
Researchers who want to make a fairer comparison end up using multiple measures and then trying, usually informally, to identify the algorithm that is most robust across those multiple measures. This is not easy to do, since there is unlikely to be a single outright winner. A sports analogy here might be that researchers try to see which recommenders perform best both when playing at home and when playing away! When a metric favours a recommender, this is like playing at home; when a metric favours other recommenders, this is like playing away. The difference is that, for soccer, for example, we have a simple and agreed way of combining the outcomes of home and away matches to obtain the points we award to each team. We have no equivalent way of combining the outcomes of different metrics.

\subsection{Measuring the relevance-diversity trade-off}

As we said before, we do not desire diversity in a top-$N$ for its own sake. We care about relevance too. Of course, there are many ways to measure the relevance of a set of recommendation. In Figure \ref{fig:rel-plots}, we show three of them: the precision, the mean reciprocal rank (MRR) and 1-call. Precision and MRR are well-known; we have more to say about 1-call in Section \ref{sec:conc}. 

This leads us to the second problem that we identified in Section \ref{sec:intro}. To identify the best algorithm requires researchers to look jointly at relevance graphs and diversity graphs.  This can be made easier, perhaps, by combining them. In the plots shown in Figure \ref{fig:tradeoff-plots}, for example, we show the three different measures of diversity (vertical axis) against precision (horizontal axis). In the plots in Figure \ref{fig:tradeoff-plots}, we are fixing $N = 10$ and we are varying $\lambda$ (Eq.\ \ref{eq:greedyobj}).

Each subfigure in Figure \ref{fig:tradeoff-plots} is divided into four  by the dotted lines that plot the precision and diversity values of the MF baseline. When, for a given value of $\lambda$, a re-ranking algorithm improves both precision and diversity over the baseline, for example, it appears as a point in the top-right quadrant. In our case study, SPAD appears to be the best of the three algorithms on this dataset: for most values of $\lambda$ it outperforms MF in both precision and diversity.

But even this is misleading! It assumes that precision and diversity are equally important. If this is not the case, then these simple plots cannot be used.

Researchers could define metrics that combine measures of relevance and diversity but this still leaves the problem of knowing how much weight this combined metric should give to each component.

With this case study, we have illustrated the problems faced by researchers who wish to compare the offline performances of recommenders that diversify their result sets. In the next section, we offer a new scoring method for comparing these algorithms.

\section{Sudden Death Score} \label{sec:sd}

%Diversification is supposed to make it more likely that the user finds an item that satisfies her. 
We introduce the Sudden Death score, a new way of comparing recommenders, whether they diversify or not: user by user, it rewards the algorithms that score hits earliest in the top-$N$.

Let $U$ be the set of users. Let $\rel_u$ be test set items that are relevant to user $u \in U$; for example, in MovieLens, $\rel_u$ can be test set items that $u$ has rated 4 or 5. 
Let $\mathit{Algs}$ be the set of recommender algorithms to be compared. Let $\RL_{u,a,N}$ be the ordered list of the top-$N$ recommendations that algorithm $a \in \mathit{Algs}$ makes to user $u \in U$. For a given user $u$, an algorithm $a$ scores a hit at position $i$ if any of the first $i$ members of $\RL_{u,a,N}$ are in $\rel_u$:
$\hit(i,u,a,N)$ is true iff $\RL\nolimits_{u,a,N}[:i] \cap \rel\nolimits_u$ is non-empty.
For user $u$, $\SD(u, a)$ is 1 iff no other algorithm in $\mathit{Algs}$ has an earlier hit.
\begin{equation}
    \SD(u, a) = \left\{ 
        \begin{array}{ll}
        1 & \mbox{if } \hit(i,u,a,N) \land \neg\exists j<i, a' \in \mathit{Algs} \hit(j,u,a',N) \\
        0 & \mbox{otherwise}
        \end{array}
                    \right.
\end{equation}
$\SD(a)$, the Sudden Death score of algorithm $a$ is the average of $\SD(u,a)$ across all users; see also Algorithm \ref{alg:sdmetric}.

The analogy with sudden death tie-breakers in sports such as badminton, fencing, golf, judo and volleyball should be clear: play stops in the tie-break period of the game as soon as one team is ahead. %Our Sudden Death score generalizes to more than two competitors and, for a given user, the `game' stops when \emph{at least} one competitor gets a hit: as soon as there is a hit, it adds 1 to the score for every competitor that has a hit at that position.

There are similarities between the Sudden Death score, which is for offline evaluation, and the framework for online evaluation described in, e.g \cite{hayes2002line}. In their framework, users choose recommendations from either a single recommendation list that has been created by interleaving the results of multiple recommender systems or from multiple recommendation lists. No actual metrics are defined in \cite{hayes2002line} but one option is to reward a system if it places the item chosen by the user earlier in its recommendation list. 

\begin{algorithm}[t]
\caption{Sudden Death}\label{alg:sdmetric}
\begin{algorithmic}[1]
\Require $U$, $\mathit{Algs}$, $N$
\Ensure $\SD(a)$ for each $a \in \mathit{Algs}$
\For {$a \in \mathit{Algs}$} $\SD(a) \gets 0$
\EndFor
\For {$u \in U$}
    \For {$a \in \mathit{Algs}$} $h_a \gets 0$
    \EndFor
    \State $i \gets 1$
    \While{$i < N$}
        \For {$a \in \mathit{Algs}$} $h_a \gets 1$ if $\hit(i,u,a,N) = \mathit{true}$
        \EndFor
        \If{$h_a = 1$ for any $a \in \mathit{Algs}$}
            \State \textbf{break}
        \EndIf
        \State $i \gets i + 1$
    \EndWhile
    \For {$a \in \mathit{Algs}$} $\SD(a) \gets \SD(a) + h_a$ 
    \EndFor
\EndFor
\State \textbf{return} $\SD(a)/|U|$ for all $a \in \mathit{Algs}$
\end{algorithmic}
\end{algorithm}

\begin{figure}[t]
\centering
\includegraphics[width=0.45\textwidth]{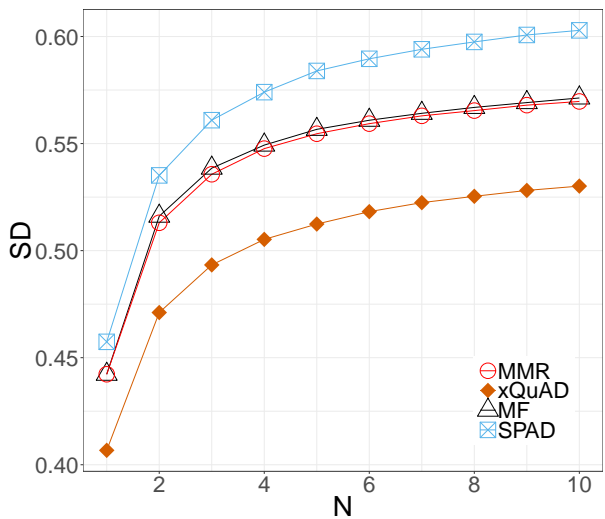}
\caption{Sudden Death scores for different values of $N$}
\label{fig:SD}
\end{figure}

Figure \ref{fig:SD} plots Sudden Death for different values for $N$ for the same algorithms we used in the previous section. In Figure \ref{fig:SD}, SPAD is the `winner' across all value of $N$. Of course, we cannot generalize from this to other datasets. For example, although we do not show the details here, in comparison with the same algorithms (MMR, xQuAD and MF), when SD scores are compared on the LibraryThing dataset, SPAD is knocked into second place by xQuAD.

\section{Concluding Remarks} \label{sec:conc}

The Sudden Death score tries to capture the idea that diversification should make it more likely that the user finds an item that satisfies her. It recognizes that diversification is not simply about creating a top-$N$ with a mix, e.g., of genres; it is about promoting relevant items into the top-$N$ to make user satisfaction more likely. 

One concern might be that it seems very similar to relevance measures such as precision and MRR, which also reward algorithms for hits and, in some cases, for earlier hits. But Figure \ref{fig:rel-plots} shows that they are different: although here SPAD is always the best, they rank MF, MMR and xQuAD differently depending on $N$. 

In fact, the Sudden Death score is perhaps most similar to Chen et al.'s 1-call metric (Figure \ref{fig:1call}), which counts how many users get at least one relevant item in the top-$N$ \cite{Chen:2006:LMP:1148170.1148245}. Although 1-call comes from Information Retrieval, Chen et al.\ also partly motivate 1-call by diversity principles: they see diversity as a means to avoid full absence of relevance, even if it means giving up some probability of achieving full relevance. The Sudden Death score can be seen as a more rank-sensitive version of principles similar to those that underlie 1-call. More specifically, the Sudden Death score and 1-call metric show different outcomes in our case study: the Sudden Death discriminates better between SPAD and the its competitors, even though the system ranking is similar overall similar (as one would also expect given the connections between the principles behind the two).

But there is a more fundamental difference too. Precision, MRR and 1-call are performance estimates. We compute them to get an estimate of how well a model will perform on future unseen data. %We can then also use these estimates to compare systems, to see which is best. 
The Sudden Death score is not a performance estimate. %It does not tell us anything about future performance. 
Its purpose is only for \emph{comparing} systems. Indeed, unlike precision, MRR and 1-call, the Sudden Death score will change depending on the algorithms we compare against. For example, xQuAD will have a particular Sudden Death score when it is compared to just SPAD but a different Sudden Death score if it is compared simultaneously to SPAD and MMR. 
It follows that the Sudden Death score does not replace precision, MRR or 1-call; it supplements them. 

\begin{acks}
This paper emanates from research supported by a grant from Science Foundation Ireland (Grant Number SFI/12/RC/2289), which is co-funded under the European Regional Development Fund.
\end{acks}

\bibliographystyle{ACM-Reference-Format}
\bibliography{sd}


\end{document}